\font\csc=cmcsc10
\input ao.sty
\hsize=210truemm \vsize=297truemm
\advance\hsize by -2truein
\advance\vsize by -2truein \advance\vsize by -\baselineskip
\overfullrule=0pt
\newdimen\ob \newif\ifdoublespace \doublespacefalse
\ob=\baselineskip
\ifdoublespace\baselineskip=1.5\baselineskip\fi
\def\ran#1#2{0\leq  #1 \leq #2}
\def\vf.{vector field}
\def\fo.{first-order}
\def\do.{differential operator}
\def\fd.{finite-dimensional}
\def\Fd.{Finite-dimensional}
\def\qes.{quasi-exactly solvable}
\def\Qes.{Quasi-exactly solvable}
\def\sc.{Schr\"odinger}

\def\all{\roq{for all}}
 \let\eps\epsilon

\def\dx{\partial_x} \def\dz{\partial_z}
\def\g{\frak g} \def\h{\frak h} \def\hn{\h^n} \def\Pn{\Cal P_n}
\def\N{\frak N} \def\Nn{\Bbb N}
\def\cp{\Bbb{CP}} \def\rp{\Bbb{RP}}
\def\hg{H_{\hbox{\sevenrm gauge}}}

\def\c{\hat c}
\def\wh{\hat \omega}
\def\Set#1#2{\bigl\{#1:#2\bigr\}}
\def\inner#1{\left\langle#1\right\rangle}
\let\union\cup
\def\asym{\mathop{\sim}\limits}
\def\abs#1{\left|#1\right|}
\catcode`\@=11
\def\Eqalign#1{\null\,\vcenter{\openup\jot\m@th
  \ialign{\strut\hfil$\displaystyle{##}$&
    $\displaystyle{{}##}$\hfil&&
    \qquad\hfil$\displaystyle{##}$&
    $\displaystyle{{}##}$\hfil\crcr
  #1\crcr}}\,}
\catcode`\@=12
\def\Ntable#1#2{\itemnumber=#1 
   \def\\{\cr \the\itemnumber.\global\advance\itemnumber by
   1&}\vcenter{\openup1\jot 
   \halign{\hfil##&&\quad $\displaystyle{}##\hfil$\cr 
   \the\itemnumber.\global\advance\itemnumber by 1 & #2\cr}}}
\def\({{\rm (}} \def\){{\rm )}}
\parskip = 2pt plus 2pt minus 2pt
\widetildes
\widehats
\barrs

\refsin
\keyin
\def\shorty#1#2{} \def\shortyear{\expandafter\shorty\the\year}
\rightline{\csc UCM--FTII Preprint
3/96}
\Title 
Quasi-Exactly Solvable Potentials on the Line\endgraf
and Orthogonal
Polynomials.
\bigskip
{\center \let\\\par \csc
Federico Finkel, Artemio Gonz\'alez-L\'opez and
Miguel A.~Rodr\'\i guez\\
\smallskip {\baselineskip=\ob\it Departamento de F\'\i sica Te\'orica II\\
Universidad Complutense\\
28040 Madrid, SPAIN\\}}
\bigskip
\centerline{March 13, 1996}
\bigskip
\Abstract In this paper we show
that a \qes. (normalizable or periodic) one-dim\-ens\-ional Hamiltonian
satisfying very mild conditions defines a family of weakly orthogonal
polynomials which obey a three-term recursion relation. In particular, we
prove that (normalizable) exactly-solvable one-dimensional systems are
characterized by the fact that their associated polynomials satisfy a two-term
recursion relation. We study the properties of the family of weakly orthogonal
polynomials defined by an arbitrary one-dimensional \qes. Hamiltonian, showing
in particular that its associated Stieltjes measure is supported on a finite
set. From this we deduce that the corresponding moment problem is determined,
and that the $k$-th moment grows like the $k$-th power of a constant as $k$
tends to infinity. We also show that the moments satisfy a constant
coefficient linear difference equation, and that this property
actually characterizes weakly orthogonal polynomial systems.
\page
\vglue .25in
\Section i Introduction.
In a recent paper, \rf{BenDun}, C.~M. Bender and G.~V. Dunne introduced a
remarkable family of orthogonal polynomials associated to the
one-dimensional Hamiltonian
$$
H = -\dx^2+\frac{(4s-1)(4s-3)}{4x^2}-(4s+4J-2)x^2+x^6,
\Eq{H}
$$
where $J$ is a positive integer and $s$ is a real parameter.
If $\psi_E(x)$ denotes an eigenfunction of $H$ with energy $E$,
the polynomials $P_k(E)$ in question are proportional to the coefficients in
the expansion of
$e^{x^4/4}x^{\frac12-2s}\psi_E(x)$ in powers of $x^2$, namely
$$
\psi_E(x) = e^{-\frac{x^4}4}x^{2s-\frac12}
\sum_{k=0}^\infty \frac{(-1)^k}{k!}
\frac{P_k(E)}{\Gamma(k+2s)}\left(\frac{x^2}4\right)^k.
$$
These polynomials are easily shown to satisfy a three-term recursion relation,
from which it follows, \rf{Chi}, that they are orthogonal with respect
to a certain Stieltjes measure $d\omega(E)$ ($\omega$ being a function of
bounded variation):
$$
\int P_k(E) P_l(E) d\omega(E)= 0,\qquad k\ne l.
\Eq{innp}
$$

The form of the coefficients of the recursion relation satisfied by the
polynomial system $\left\{P_k(E)\right\}_{k=0}^\infty$ implies
that this system has several remarkable properties. First
of all, the norm of the polynomials $P_k$ with $k\ge J$ vanishes. Thus, the
polynomials $P_k$ form what is called a {\it weakly orthogonal}
polynomial system,
\rf{Chi}. To be precise, we shall use from now on the term {\it orthogonal
polynomial system} for a family of orthogonal polynomials
$\left\{P_k(E)\right\}_{k=0}^\infty$ with $\deg P_k=k$ for all
$k$, and such that the norm of $P_k$ does not vanish for any $k$.

Secondly, each
$P_k$ with
$k\ge J$ factors into the product of
$P_J$ and another polynomial,
\ie
$$
P_{J+m} = P_J\, Q_m, \qquad m\ge 0,
$$
where $Q_m$ has degree $m$. Finally, the $J$ simple real
zeros of $P_J$ are eigenvalues of $H$ whose corresponding eigenfunctions,
being the product of the factor 
$$\mu(x) = e^{-\frac{x^4}4}x^{2s-\frac12}$$
times a polynomial in $x^2$, are square-integrable. The existence of these
exactly computable eigenfunctions and eigenvalues of $H$ had been deduced
before, \rf{Tur}, \rf{GKOnorm}, from the fact that $H$ is {\it \qes.,}
meaning that it is an element of the enveloping algebra of a certain
realization of
$\sL{2,\R}$ in terms of first-order differential operators acting on a
finite-dimensional subspace of the space of $C^\infty$ functions (see the
next section for more details).
The above results strongly suggest, \rf{KUW}, that
there is a connection between \qes. Hamiltonians and certain
families of weakly orthogonal polynomials. In this paper, we show in
detail that this is indeed the case for all one-dimensional \qes.
Hamiltonians, both normalizable and periodic, satisfying very general
conditions. The paper is organized as follows.

Using the results on quasi-exact solvability reviewed in
\section{qes}, we explain in \section{rr} how to construct the weakly
orthogonal polynomial system associated to each of the normal forms
of a one-dimensional \qes. Hamiltonian listed in \rf{GKOnorm}, \rf{GKO}.
Like the polynomial system introduced in \rf{BenDun}, this system
always satisfies a three-term recursion relation, whose coefficients we
explicitly compute. This allows us to prove that one-dimensional
(normalizable) exactly solvable Hamiltonians are characterized by the fact that
their associated polynomials satisfy a two-term recursion relation.
In \section{op} we show that the polynomials associated to
an arbitrary one-dimensional \qes. Hamiltonian enjoy properties completely
akin to those listed above for the Hamiltonian \eq{H}. We also study
in this section the properties of the moment functional defined by the family
of weakly orthogonal polynomials of a \qes. Hamiltonian, giving a rigorous
proof of the fact that its associated Stieltjes measure is supported on a
finite set, \rf{KUW}, so that the integral \eq{innp} reduces to a finite sum. 
From this we deduce that the associated (Hamburger or Stieltjes) moment
problem is determined, and that the $k$-th moment behaves like the $k$-th
power of a constant for large $k$, illustrating this statement with an explicit
example for the Hamiltonian \eq{H}. We also show that the moments satisfy a
constant coefficient linear difference equation, a property which
in fact characterizes weakly orthogonal polynomial systems. The paper ends
(\section{conc}) with a brief review of these results, stressing the role
played by weak orthogonality---as opposed to true orthogonality---in their
derivation.

\Section{qes} Quasi-exactly Solvable Potentials.
For the reader's convenience, we present in this Section a summary of
the major results in the theory of \qes. systems that we shall need in the
sequel. A one-dimensional
\sc. operator (or Hamiltonian)
$H = -\dx^2+V(x)$ is {\it \qes.} if there exists a \fd. Lie algebra of
first-order
\do.s
$$
\g = \Span\set{\xi_a(x)\dx+\eta_a(x)}{1\le a\le r}\equiv
\Span\set{T_a(x)}{1\le a\le r}
$$
such that:
\item{i)}$\g$ leaves invariant a \fd. module of smooth
functions $\N\subset\C^\infty(\R)$, \ie
$X\cdot f\in\N$ for all $f\in\N$ and all $X\in\g$. In other words, $\g$ admits
a finite-dimensional representation in terms of smooth functions.
\item{ii)}$H$ is in the universal enveloping algebra of $\g$, \ie $H$ can be
expressed as a polynomial in the
generators $T_a$, $\ran ar$, of $\g$.

A Lie algebra of first-order \do.s satisfying i) is
called {\it \qes..} A Hamiltonian $H$ satisfying condition ii) above for an
arbitrary (not necessarily \qes.) Lie algebra $\g$ is said to be {\it
Lie-algebraic.}

\smallskip
If $H$ is \qes., it follows that the restriction of $H$ to $\N$ is a
\fd. linear operator $\N\to\N$, and therefore the eigenfunctions of $H$
lying in $\N$ and its corresponding eigenvalues can be exactly computed by
purely algebraic methods (diagonalizing a square matrix of order
$\dim\N$). We shall refer to these
eigenfunctions of
$H$ as lying in $\N$ as its {\it algebraic eigenfunctions} (although, of
course, they need not be algebraic functions in the technical sense of the
word). The functions in $\N$
need not a priori satisfy any boundary
conditions (like square-integrability, periodicity, vanishing at
the endpoints, etc.) coming from the physics of the problem,
whose mathematical purpose is to guarantee that $H$ is a self-adjoint
operator. If they do, then the restriction of $H$ to $\N$ is self-adjoint, and
therefore $H$ has exactly
$\dim\N$ linearly independent algebraic eigenfunctions, whose corresponding
$\dim\N$ {\it real} eigenvalues (counting multiplicities) are
exactly (\ie algebraically) computable. We shall say in this case that the
\qes. potential $H$ (or the potential $V$) is {\it fully algebraic}. See
\rf{GKOnorm} and \rf{GKO} for an in-depth discussion of fully algebraic
potentials under the boundary condition of square-integrability on $\R$.

It can be shown (\cf \rf{GKOqes}) that a \qes. \sc. operator $H$ can be
expressed as a polynomial of degree at most two in the generators $T_a$, $\ran
ar$, of $\g$. Moreover, a well known theorem, \rf{Tur}, \rf{KOlado},
\rf{GKOqes}, asserts that every \qes. Lie algebra of first-order \do.s $\g$
is related by a (local) change of variable
$$
z=\zeta(x) \Eq{cv}
$$
and a {\it gauge transformation} with gauge factor $\mu(z)>0$ to (a subalgebra
of) one of the Lie algebras
$\g^n=\h^n\oplus\R$, where
$\h^n=\Span\{J_-^n,J_0^n,J_+^n\}\approx\sL{2,\R}$,
$$
J_-^n = \dz,\qquad J_0^n = z\dz-\frac n2,\qquad J_+^n=z^2\dz-n\,z
\Eq{jeps}
$$
and $n$ is a nonnegative integer. In other
words, every element
$X(x)\in\g$ is of the form
$$
X(x) = \left.\mu(z)\cdot J(z)\cdot
\frac1{\mu(z)}\right|_{z=\zeta(x)},\qquad J(z)\in\g^n,
$$
for some fixed $n$. This implies that the {\it gauge
Hamiltonian}
$$
\hg(z) = \left.\frac1{\mu(z)}\cdot H(x)\cdot \mu(z)\right|_{x=\zeta^{-1}(z)}
\Eq{hgh}
$$
is also a polynomial of degree at most two in the generators $J_\eps^n$,
\ie (dropping the explicit $n$ dependence in the generators $J_\eps^n$)
$$
-\hg=\sum_{a,b} c_{ab}\,J_a\,J_b+\sum_a c_a\,J_a + c_*,
\Eq{hgj}
$$
for some real constants $c_*$, $c_a$, and $c_{ab}=c_{ba}$ (the minus sign is
for later convenience). The spectral problems of $H$ and
$\hg$ are related in an obvious way: indeed, from \eq{hgh} it follows that if
$\chi(z)$ is an eigenfunction of $\hg$ with eigenvalue $E$ then
$$
\psi(x) = \left.\mu(z) \chi(z)\vphantom{_p}\right|_{z=\zeta(x)}
\Eq{algwf}
$$ will be an eigenfunction of $H$ with the same eigenvalue (not taking into
account the boundary conditions). Since the Lie algebra $\g^n$
admits as invariant module the space $\CP_n$ of real polynomials of degree at
most
$n$ in $z$, if $H$ is fully algebraic then $\hg$ has $n+1$ linearly
independent algebraic eigenfunctions lying in
$\CP_n$. Hence
$H$ has $n+1$ linearly
independent algebraic eigenfunctions of the form \eq{algwf}, with
$\chi\in\CP_n$ a polynomial of degree at most $n$.

From \eq{jeps} and \eq{hgj} it follows, \rf{GKOnorm}, that the gauge
Hamiltonian is of the form\goodbreak
$$
\Eeqn{
-\hg &= P(z)\,\dz^2+\left\{Q(z)-\frac{n-1}2P'(z)\right\}\,\dz\\
&\kern 1in +\left\{R-\frac n2 Q'(z)+\frac{n(n-1)}{12}P''(z)\right\},
\Eq{hgz}}
$$
where $P$, $Q$ and $R$ are polynomials of degrees
4, 2 and 0, respectively, given by
$$\eeqn{
P(z) = c_{++}z^4+2c_{+0}z^3+c_{00}z^2+2c_{0-}z+c_{--},\Eq P\\
Q(z) = c_ + z^2+c_0 z + c_-,\Eq Q\\
R = \frac{n(n+2)}{12}c_{00}+c_*.\Eq R
}
$$
Note that, due to the Casimir relation
$$
J_0^2-\frac12(J_+J_-+J_-J_+) = \frac n4(n+2),
$$
we have set, without loss of generality, $c_{+-}=0$. There are also explicit
formulas for the change of variables \eq{cv} and gauge factor
$\mu(z)$ needed to put the differential operator \eq{hgz} in \sc. form, \cf
\rf{GKOnorm}. Indeed, assuming that $P(z)>0$ on an interval $I$ then for
$z\in I$ we have
$$
x = \zeta^{-1}(z) = \int^z\frac{dy}{\sqrt{P(y)}},\qquad
\mu(z) = P(z)^{-n/4}\exp\left\{\int^z\frac{Q(y)}{2P(y)}dy\right\}
\Eq{xmu}
$$
and
$$
V(x) = -R+\left.\frac{-n(n+2)\left (P P'' -
\fr34 P'^2\right ) - 3(n+1) \left( Q P' - 2 P
Q'\right) +  3Q^2}{12P}\right|_{z=\zeta(x)},
\Eq{pot}
$$
where the primes denote derivatives with respect to $z$.

The canonical form \eq{hgz} of the \qes. Hamiltonian
$H$ is not unique, since there is a residual symmetry group preserving the Lie
algebra $\hn$, given by the adjoint action on
$\hn$ of the Lie group of transformations generated by $\g^n=\hn\oplus\R$.
More precisely, the elements of $\g^n$ are the infinitesimal generators of the
standard $\GL{2,\R}$ action on the space $\Pn$, given by
$$
p(z)\in\Pn\mapsto \hat p(w) = (\gamma w+\delta)^n p\left(\frac{\alpha
w+\beta}{\gamma w + \delta}\right),\qquad
\pmatrix{\alpha &\beta\cr\gamma &\delta}\in \GL{2,\R}.
\Eq{rhon}
$$
We shall denote, as is customary, by $\rho_{n}$ this (irreducible)
multiplier representation of $\GL{2,\R}$ on $\Pn$.
Note that the action \eq{rhon} is just the composition of the projective
transformation
$$
z = \frac{\alpha w+\beta}{\gamma w+\delta}
$$
and the gauge transformation with gauge factor $\mu(w) = (\gamma w+\delta)^n$.
The adjoint action of
$\GL{2,\R}$ on $\hn$ induced by \eq{rhon} is given by
$$
J(z)\mapsto 
\Jhat(w) = (\gamma w + \delta)^n\cdot
J\left(\frac{\alpha w+\beta}{\gamma w + \delta}\right)\cdot
(\gamma w + \delta)^{-n}.
\Eq{jt}
$$
A straightforward calculation, \rf{GKOnorm}, shows that the generators of
$\hn$ transform under the representation
$\rho_{2,-1}$---where $\rho_{n,i}=\rho_n\otimes\det^i$,
$\det:A\mapsto\det A$ being the standard determinantal representation---
independently of $n$. As a consequence of all this, the transformed
differential operator
$$
\widehat\hg = (\gamma w + \delta)^n\cdot \hg\left(\frac{\alpha w+\beta}{\gamma
w + \delta}\right)\cdot (\gamma w + \delta)^{-n}
\Eq{hghat}
$$
is still of the form \eq{hgz}, with $P$, $Q$ and $R$ replaced by
appropriate polynomials $\Phat$,
$\Qhat$ and $\Rhat$ of respective degrees 4, 2 and 0. It can be shown,
\cf\rf{GKOnorm}, that $\Rhat=R$ and
$$
\Phat(w) = \frac{(\gamma w+\delta)^4}{\Delta^2}
P\left(\frac{\alpha w+\beta}{\gamma w + \delta}\right),\qquad
\Qhat(w) = \frac{(\gamma w+\delta)^2}\Delta Q\left(\frac{\alpha
w+\beta}{\gamma w + \delta}\right),
\Eq{PQ}
$$
with
$$
\Delta=\det\pmatrix{\alpha&\beta\cr\gamma&\delta}.
$$
Hence the polynomials $P$, $Q$ and $R$ determining the
differential operator $\hg$ transform under the representations
$\rho_{4,-2}$, $\rho_{2,-1}$ and $\rho_{0}$ of $\GL{2,\R}$. Furthermore,
the algebraic eigenfunctions of $\hg$ clearly transform under the
representation $\rho_n$; indeed, if $\chi(z)$ is
an eigenfunction of
$\hg$ with eigenvalue $E$ then it follows from \eq{hghat} that 
$$
\hat\chi(w) = (\gamma w + \delta)^n\cdot \chi\left(\frac{\alpha
w+\beta}{\gamma w + \delta}\right)
\Eq{chih}
$$
is an eigenfunction of $\widehat\hg$ with the same eigenvalue.

In \rf{GKO} and \rf{GKOnorm}, the form-invariance of the differential operator
$\hg$ under the $\GL{2,\R}$ action \eq{hghat} described above was exploited to
place
$\hg$ in canonical form. Indeed, it can be shown that there are ten
inequivalent real normal forms for a (nonzero) fourth-degree polynomial
$P$ \rf{fnote1}
transforming under the representation $\rho_{4,-2}$ of $\GL{2,\R}$, each of
which leads to a canonical form for $\hg$. Of these ten canonical forms, five
correspond to {\it normalizable} Hamiltonians, whose algebraic eigenfunctions
are square-integrable (provided the coefficients $c_{ab}$ and $c_a$ satisfy
certain inequalities), and the remaining are associated to Hamiltonians
with periodic potentials. The five normal forms associated to normalizable
Hamiltonians, which are characterized by the fact that $P$ has at least one
multiple root on the real projective line $\rp$, are given by%
%
%
$$\ntable{
\nu(z^2+1),\\
\nu(z^2-1),\hphantom{\bigl(1-k^2(1-z^2)\bigr)}\\
\nu z^2,\\
z,\\
1,}\Eq{normlist}
$$
where $\nu>0$ is a real parameter. For example, the \qes. potential discussed in
\rf{BenDun} corresponds to the fourth normalizable canonical form $P(z)=z$.
The remaining normal forms,
corresponding to periodic potentials, are
$$
\Ntable6{
\nu(1-z^2)(1-\kappa^2 z^2),\\
\nu(1-z^2)\bigl(1-\kappa^2(1-z^2)\bigr),\\
\nu(1+z^2)\bigl(1+(1-\kappa^2) z^2\bigr),\\
\nu(1+z^2)^2,\\
\nu(1-z^2),
}\Eq{perlist}
$$
where $\nu>0$, $0<\kappa<1$.

\Section{rr} The Recursion Relation.
Let $H=-\dx^2 +V(x)$ be a \qes. Hamiltonian. From the previous section, we
know that there is a change of variable \eq{cv} and gauge factor $\mu(z)>0$
such that
$H(x)=\left.\mu(z)\cdot\hg(z)\cdot\frac1{\mu(z)}\right|_{z=\zeta(x)}$,
with $\hg$ given by \eq{hgj} (and $c_{+-}=0$). Furthermore, if $H$ is fully
algebraic then it has
$n+1$ algebraic eigenfunctions of the form \eq{algwf}, with
$\chi(z)\in\CP_n$ an eigenfunction of $\hg$. Let $\chi_E(z)$ be
an eigenfunction of $\hg$ with eigenvalue $E$ (not necessarily a polynomial in
$z$). Writing
$$
\chi_E(z) = \sum_{k=0}^\infty P_k(E)\chi_k(z),
\Eq{chie}
$$
where
$$
\chi_k(z) = \frac{z^k}{k!},\qquad k\ge0,
$$
and taking into account that
$$
J_-\cdot\chi_k = \chi_{k-1},\qquad
J_0\cdot\chi_k = \left(k-\frac n2\right)\chi_k,\qquad
J_+\cdot\chi_k = (k-n)(k+1)\chi_{k+1},
\Eq{jaction}
$$
\cf \eq{jeps}, we easily find that the coefficients $P_k(E)$ satisfy the
following five-term recursion relation:
\goodbreak
$$
\Eeqn{
-c_{--}P_{k+2} &= 
\left[(2k-n+1)c_{0-}+c_-\right]P_{k+1}\\&\quad{} +
\left[E+c_*+c_0\bigl(k-\frac n2\bigr)+c_{00}\bigl(k-\frac
n2\bigr)^2\right]P_{k}\\
&\quad {}+k(k-1-n)\left[(2k-n-1)c_{+0}+c_+\right]P_{k-1}\\&\quad{} 
+k(k-1)(k-1-n)(k-2-n)\,c_{++}P_{k-2},\qquad
k\ge 0.
\Eq{grr}}
$$

If $c_{--}\ne0$, the general solution of the recursion relation \eq{grr}
depends on the two arbitrary functions $P_0(E)$ and $P_1(E)$. This
simply reflects the fact that when $c_{--}\ne0$ the leading coefficient
$P(z)$ of
$\hg$ does not vanish at $z=0$ (\cf\eq P); thus, the differential equation
$(\hg-E)\,\chi_E = 0$ has a regular point at the origin, and therefore it
admits two linearly independent solutions \eq{chie} analytic at 0. If
$P_0(E)$ and $P_1(E)$ are chosen to be polynomials in $E$, then \eq{grr}
implies that all the coefficients $P_k(E)$ are polynomials in $E$. However,
the general recursion relation \eq{grr} suffers from two major drawbacks.
In the first place, even if we choose $P_0(E)$ and $P_1(E)$ as polynomials of
degree 0 and 1 in $E$, respectively, \eq{grr} is incompatible with the
desirable property that $P_k(E)$ be of degree $k$ in $E$ for all $k$,
unless $c_{--}=0$. Secondly, even in this  case \eq{grr} will be in general a
four-term recursion relation, implying that the polynomials $P_k(E)$ may not
be orthogonal with respect to any (nonzero) Stieltjes measure $d\omega(E)$.
Indeed, it is well known,
\rf{Chi}, \rf{Erd}, that a necessary and sufficient condition for a
family of polynomials $\left\{P_k\right\}_{k=0}^\infty$ (with $\deg P_k=k$)
to form an orthogonal polynomial system is that $P_k$ satisfies a {\it
three\/}-term recursion relation of the form
$$
P_{k}=(A_k E + B_k) P_{k-1} + C_{k} P_{k-2},\qquad k\ge1,
\Eq{rr}
$$
where the coefficients $A_k,B_k,C_k$ are {\it independent of $E$,}
$A_k\ne0$, $C_1=0$, and $C_k\ne0$ for $k\ge1$. If the coefficient $C_k$ in
\eq{rr} vanishes for some positive integer $k$, then this recursion relation
only defines a {\it weakly} orthogonal polynomial system \rf{fnote2}.
It is one of the main goals of this paper to show that
both difficulties described above can always be overcome, provided (roughly
speaking) that we expand the eigenfunction $\chi_E$ with respect to an
appropriate variable. This will be achieved by using the non-uniqueness of
$\hg$, due to the $\GL2$ symmetry described in the previous section, to place
$\hg$ in a suitable canonical form.

From the form of the recursion relation \eq{grr},
it follows that both difficulties described above
disappear if
$$
c_{--} = c_{++} = 0.
\Eq{cond}
$$
Indeed, if \eq{cond} holds then \eq{grr} reduces to the three-term recursion
relation
$$
\seqn{
-\left[(2k-n-1)c_{0-}+c_-\right]P_{k}=\\
\kern3em\left[E+c_*+c_0\bigl(k-\frac n2-1\bigr)+c_{00}\bigl(k-\frac
n2-1\bigr)^2\right]P_{k-1}\\
\kern3em{}+(k-1)(k-2-n)\left[(2k-n-3)c_{+0}+c_+\right]P_{k-2},
\qquad k\ge 1,
\Eq{trr}}
$$
which uniquely determines all the functions $P_k(E)$ in terms of $P_0(E)$
provided that, for all positive integer values of $k$, the coefficient of the
left-hand side of
\eq{trr} does not vanish. If $P_0(E)$ is taken as a constant, for instance if
$P_0(E)=1$, then \eq{trr} implies that $P_k(E)$ is a polynomial of degree $k$
in $E$ for all $k\ge0$.

Let us see now that we can always arrange for \eq{cond}
to be satisfied, by using the action \eq{PQ} to transform $P(z)$ into a
normal form $\Phat(w)$ for which \eq{cond} holds. Indeed,
\eq{cond} simply states that the polynomial
$P(z)$ vanishes at $z=0$ and $z=\infty$, when $z$ is allowed to vary over
the complex projective line $\cp$. Note that we need $z$ to belong to the
{\it complex} projective line at this stage so that $P$ is
guaranteed to have a root, which is essential for the argument that follows.
Consequently, the $\GL{2,\R}$ action described in the previous section will be
replaced in what follows by a $\GL{2,\C}$ action.

We can assume, first of all, that $P(z)$ is one of the normal forms listed
in equations \eq{normlist} and \eq{perlist}. We must distinguish three cases,
characterized by the position of the roots of
$P$ in the complex projective line. Indeed, either $P$ has two different
roots $z_1\ne z_2$ in
$\cp$, or it has four coincident roots. In the first case, either one of the
roots is at infinity, or both roots are finite.

{\it Case 1:} $P$ has two different roots $z_1\ne z_2=\infty$.%
\par\noindent
This case occurs when $P$ is 
one of the first four normalizable canonical forms \eq{normlist}, or the fifth
periodic canonical form \eq{perlist}. In this case, the translation $w=z-z_1$
transforms $P(z)$ into a  polynomial
$\Phat(w)$ vanishing at zero and infinity. In the original $z$ coordinate,
by \eq{chih} this amounts to replacing \eq{chie} by
$$
\chi_E(z) = \sum_{k=0}^\infty P_k(E)\frac{(z-z_1)^k}{k!}.
\Eq{alt1}
$$
In other words, we expand $\chi_E(z)$ as a power series around
the point $z=z_1$, which is a singular point of the linear differential
equation $(\hg-E)\chi_E=0$ (if $z_1$ is
a simple root of $P$, $z_1$ is actually a regular singular point, whose
indicial equation is easily seen to have 0 as a root).  By
\eq{algwf}, in the ``physical" coordinate
$x$
\eq{alt1} becomes
$$
\psi_E(x) = \mu\bigl(\zeta(x)\bigr)\, \sum_{k=0}^\infty
P_k(E)\frac{(\zeta(x)-z_1)^k}{k!}.
\Eq{xalt1}
$$

{\it Case 2:} $P$ has two different finite roots $z_1\ne z_2$.%
\par\noindent
This is the case when $P$ is one of the first four periodic normal
forms \eq{perlist}. The projective transformation
$w=(z-z_1)/(z-z_2)$ will again transform 
$P(z)$ into a polynomial $\Phat(w)$ vanishing at $w=0,\infty$. Going back
to the original $z$ coordinate, by \eq{chih} we just have to replace \eq{chie}
by
$$
\chi_E(z) = (z-z_2)^n\,\sum_{k=0}^\infty\frac1{k!}
P_k(E)\left(\frac{z-z_1}{z-z_2}\right)^k,
\Eq{alt2}
$$
apart from an inessential overall factor. In terms of the physical coordinate
$x$, \eq{alt2} can be written as
$$
\psi_E(x) = \mu\bigl(\zeta(x)\bigr)\,
(\zeta(x)-z_2)^n\,\sum_{k=0}^\infty\frac1{k!}
P_k(E)\left(\frac{\zeta(x)-z_1}{\zeta(x)-z_2}\right)^k.
\Eq{xalt2}
$$

{\it Case 3:} $P$ has a quadruple root.%
\par\noindent
This corresponds to the fifth normalizable canonical form, $P=1$, which has a
quadruple root at infinity. Note that $P=1$ implies that the physical
coordinate $x$ can be taken as the canonical coordinate $z$.
By \eq{P}, we have
$$
c_{++}=c_{+0}=c_{00}=c_{0-}=0,\qquad c_{--}=1.
$$
Performing an additional translation, if necessary,
we can also take without loss of generality $c_-=Q(0)=0$ (notice that $P$
is constant, and therefore does not change under translations). Thus equation
\eq{grr} reduces in this case to
$$
-P_{k+2} =
\left[E+c_*+c_0\bigl(k-\frac n2\bigr)\right]P_{k}
+k(k-1-n) c_+ P_{k-1},\qquad
k\ge 0.
\Eq{ghr}
$$
Since $P=1$ is the fifth normalizable case of references
\rf{GKOnorm},
\rf{GKO}, $c_+$ must vanish if we want $H$ to be {\it normalizable}, \ie the
algebraic eigenfunctions of
$H$ to be square-integrable. Therefore, in this case \eq{ghr} reduces to
$$
-P_{k+2} =
\left[E+c_*+c_0\bigl(k-\frac n2\bigr)\right]P_{k},\qquad
k\ge 0,
$$
which is equivalent to two two-term recursion relations for the even and odd
coefficients $P^0_j=P_{2j}$ and $P^1_j=P_{2j+1}$, namely
$$
-P^\eps_{j+1} =
\left[E+c_*+c_0\bigl(2j+\eps-\frac n2\bigr)\right]P^\eps_{j},\qquad j\ge 0;
\quad\eps=0,1.
\Eq{reo}
$$
Note that in this case the potential is $V(x)=\frac14c_0^2x^2-c_*$ (with
$c_0<0$), \cf \rf{GKOnorm}.

To complete the discussion of Cases 1 and 2, we still have to deal with an
important technical issue; namely, we must find under what conditions the
coefficient of
$P_k$ in \eq{trr} never vanishes for positive integer values of $k$. Let
$\Phat$ and $\Qhat$ be the transforms of
$P$ and $Q$ under the projective transformation $z\mapsto w$ defined in the
foregoing discussion of Cases 1 and 2; note that, by construction, $\Phat(w)$
vanishes at
$w=0,\infty$. The coefficient of interest can be
expressed as
$$
(2k-n-1)\,\c_{0-}+\c_-,\qquad k\ge1,
\Eq{ccoeff}
$$
where
$$
\c_{0-}=\frac12\Phat'(0),\qquad \c_-=\Qhat(0).
$$
From \eq{PQ} it easily follows that
$$
\c_{0-}=\frac12 P'(z_1),\qquad \c_-=Q(z_1)
\Eq{ctrans}
$$
for Case 1 ($w=z-z_1$), and
$$
\c_{0-}=\frac{P'(z_1)}{2(z_1-z_2)},\qquad \c_-=\frac{Q(z_1)}{z_1-z_2}
\Eq{cproj}
$$
for Case 2 ($w=(z-z_1)/(z-z_2)$). We shall now distinguish three subcases:

{\it Case i.} $z_1$ is a simple real root of $P$%
\par\noindent
This case occurs when $P$ is one of the canonical forms 2, 4, 6, 7, or 10.
Note that in this case the mapping $z\mapsto w$ is real, and so are
the coefficients $\c_{0-}$, $\c_-$. From \eq{xmu} and
\eq{ctrans}--\eq{cproj} it is immediate to deduce the asymptotic formulas
$$x \asym_{z\to z_1} \abs{z-z_1}^{\frac12},\qquad
\mu(z) \asym_{z\to z_1}
\abs{z-z_1}^{\frac14\left(\frac{\c_-}{\c_{0-}}-n\right)},
$$
where we have dropped unessential
constant multiplicative factors from the right-hand side, and have taken for
convenience $z_1$ as the lower limit of the integral giving $x$ in terms
of $z$. We saw in the previous section that when $H$ is fully algebraic it has
$n+1$ linearly independent algebraic eigenfunctions of the form \eq{algwf},
where $\chi\in\Pn$. It follows that the polynomial factor
$\chi(z)$ cannot vanish at the origin for all the algebraic eigenfunctions
of $H$. Hence there is at least one algebraic eigenfunction of $H$ whose
asymptotic behavior at $x=0$ is given by
$$
\psi(x)\asym_{x\to0}\abs x^{\frac12\left(\frac{\c_-}{\c_{0-}}-n\right)}.
$$
If {\it all} the algebraic eigenfunctions of $H$ are regular at
$x=0$, then we must have
$$
\frac{\c_-}{\c_{0-}}-n\ge 0.
\Eq{cineq}
$$
Since \eq{ccoeff} can
be written as
$$
2\c_{0-}\left[ \frac12\left(\frac{\c_-}{\c_{0-}}-n\right) +
\left(k-\frac12\right)\right],
$$
it follows from \eq{cineq} that the coefficient \eq{ccoeff} cannot vanish in
this case.

{\it Case ii.} $z_1$ is a simple complex root of $P$%
\par\noindent
In this case $P$ is either the first or the eighth canonical form. Since $z_1$
is not real, the mapping $z\mapsto w$ is not real either, and the above
asymptotic argument is not valid (the eigenfunctions of $H$ need not be
regular outside the real axis).
For the first canonical form \eq{normlist}, we can take $w=z-i$ and therefore
$$
\c_{0-} = i\,\nu,\qquad \c_- = c_- - c_+ +i\,c_0
$$
from \eq{ctrans}. Hence the coefficient \eq{ccoeff} does not
vanish in this case provided that the following conditions are satisfied:
$$
c_-\ne c_+ \qquad\hbox{\rm or}\qquad
\frac12\left(n+1-\frac{c_0}\nu\right)\ne 1,2,\dots.
\Eq{c1}
$$
It is easily checked that the choice $w=z+i$ leads exactly to the same
conditions.
For the eighth canonical form, we can take $w=(z-i)/(z+i)$, and
therefore, from \eq{cproj},
$$
\c_{0-} = \frac12\nu \kappa^2,\qquad \c_- = \frac{c_0}2+\frac i2(c_+-c_-).
$$
Hence in this case the conditions for the coefficient \eq{ccoeff} not to
vanish are
$$
c_-\ne c_+ \qquad\hbox{\rm or}\qquad
\frac12\left(n+1-\frac{c_0}{\nu\kappa^2}\right)\ne 1,2,\dots.
\Eq{c8}
$$
It is straightforward to check that the choice $w=(z+i)/(z-i)$ yields
the same conditions, while the other natural choice
$w=(\sqrt{1-\kappa^2}z\mp i)/(\sqrt{1-\kappa^2}z\pm i)$ only has the effect of
replacing the first condition
\eq{c8} by
$c_+\ne(1-\kappa^2)c_-$.

{\it Case iii.} $z_1$ is a multiple root of $P$%
\par\noindent
This case takes place when $P$ is either the third or the ninth canonical
form, and in both cases \eq{ccoeff} reduces to $\c_-$.
For the third canonical form \eq{normlist}, if $c_-\ne0$ then we take $w=z$,
and therefore $\c_-=c_-\ne0$. If
$c_-=0$, then $c_+\ne0$ if all the algebraic eigenfunctions of $H$ are
square-integrable (see \rf{GKOnorm}). Hence, taking $w=1/z$, we get
$\Phat(w)=\nu w^2$ and $\Qhat=-(c_++c_0 w)$, so that $\c_{0-}=0$ and $\c_- =
-c_+\ne0$. Hence the coefficient \eq{ccoeff} cannot vanish in this case.
Finally, if $P$ is the ninth canonical form \eq{perlist} then $w=(z-i)/(z+i)$
and
$$
\c_- = \frac{c_0}2+\frac i2(c_+-c_-).
$$
Hence \eq{ccoeff} will not vanish if
$$
c_+\ne c_-\qquad\hbox{\rm or}\qquad c_0\ne 0.
\Eq{c9}
$$
Note that when \eq{c9} does not hold $V$ reduces to a constant potential:
$$
V = \frac{c_-^2}{4\nu}-\frac5{12} n(n+2)-c_*.
$$

In summary, the previous analysis shows that the critical coefficient
\eq{ccoeff} cannot vanish for any positive integer $k$ provided that $V$ is
fully algebraic, that all its algebraic eigenfunctions are
regular (or square-integrable, for the third normalizable canonical form
\eq{normlist}), and that conditions \eq{c1},
\eq{c8}, and \eq{c9} are satisfied when $P$ is one of the normal forms 1, 8
or 9, respectively. If
\eq{ccoeff} doesn't vanish, defining new polynomials
$\Phat_k$ by
$$
P_k = \cases{\displaystyle
\frac{(-1)^k}{(2\c_{0-})^k}\,
\frac{\Phat_k}{\Gamma\left(\frac{\c_-}{2\c_{0-}}+k-\frac n2+\frac12\right)},&
\qquad if $\c_{0-}\ne0$;\cr\displaystyle
\frac{(-1)^k}{\c_-^k}\,\Phat_k,&\qquad if $\c_{0-}=0$
}
\Eq{phatk}
$$
the recursion relation \eq{trr} can be written in the more standard form
$$
\seqn{
\Phat_{k+1} =
\left[E+c_*+\c_0\left(k-\frac n2\right)+\c_{00}\left(k-\frac
n2\right)^2\right]\,\Phat_k\\\quad-
k(k-n-1)\left[\c_{+0}(2k-n-1)+\c_+\right]
\left[\c_{0-}(2k-n-1)+\c_-\right]\Phat_{k-1},
\quad k\ge0.
\Eq{nrr}
}
$$
We have thus proved the main theorem in this section:
\Th{main}
Let\/ $V$ be a fully algebraic one-dimensional \qes. potential
whose algebraic eigenfunctions are all regular \(or
normalizable, if\/ $V$ corresponds to the third or fifth  canonical forms
in \eq{normlist}\). Assume, furthermore, that
conditions \eq{c1}, \eq{c8} or \eq{c9} are satisfied, if\/ $V$ is obtained from
the first, eighth or ninth canonical forms \eq{normlist}--\eq{perlist},
respectively. Then\/ $V$ defines a family of weakly orthogonal polynomials
$\bigl\{\Phat_k\bigr\}_{k=0}^\infty$ satisfying a three-term recursion
relation \eq{nrr} \(or \eq{reo}, if\/ $V$ corresponds to the fifth
canonical form\). The polynomials $\Phat_k$ are defined by
\eq{phatk} and \eq{xalt1}, if\/ $V$ is associated to one of the
canonical forms 1--4 or 10,
or by \eq{phatk} and \eq{xalt2}, if\/ $V$ corresponds to one of the normal
forms 6--9. Finally, the potential\/
$V$ associated to the fifth  canonical form
defines two families of weakly orthogonal polynomials $P^0_j=P_{2j}$ and
$P^1_j=P_{2j+1}$ through \eq{xalt1}.

We shall say that a \qes. potential $V$ is {\it exactly solvable} if it is
independent of the ``spin" parameter $n$. This implies that $V$ has $n$
algebraic eigenvalues and eigenfunctions for arbitrary $n\in\Nn$, so that we
can algebraically compute an infinite number of eigenvalues of $V$ (leaving
aside the boundary conditions). All exactly solvable normalizable
one-dimensional potentials have been classified; see
\rf{GKOnorm} for a complete list. The quintessential example of exactly
solvable one-dimensional potential is the harmonic oscillator potential, which
corresponds to the fifth canonical form \eq{normlist}. We have seen in the
previous section that in this case there are two families of orthogonal
polynomials (the odd and even coefficients in \eq{xalt1}), each of which
satisfies a two-term recursion relation \eq{reo}. 
We shall now show that, as conjectured in \rf{BenDun}, the latter property
actually characterizes exactly solvable normalizable potentials:

\Th{es}
The weakly orthogonal polynomial system associated to an exactly
solvable normalizable potential satisfies
a two-term recursion relation.

\Proof
The proof is a simple case-by-case analysis using the classification
of exactly solvable normalizable potentials given in \rf{GKOnorm}.
Indeed, for the first normalizable canonical form $P(z)=\nu(z^2+1)$ we have
$w=z\mp i$, and therefore $\Phat(w) = P(w\pm i) = \nu w(w\pm 2i)$, so that
$\c_{+0}=0$. Since $\Qhat(w) = Q(w\pm i)$, we also have $\c_+ = \Qhat''(0)/2 =
Q''(\pm i)/2 = c_+$. But the exactly solvable potentials associated to this
normal form are characterized by the vanishing of $c_+$, \rf{GKOnorm}, so that
$\c_+=\c_{+0}=0$, and \eq{nrr} is a two-term recursion relation.
Similarly, for the second normalizable canonical form, $P(z)=\nu(z^2-1)$ and,
for instance, $w=z\mp1$. Proceeding as before we obtain that $\c_{+0}=0$ and
$\c_+=c_+$. Since exactly solvable potentials are again those satisfying the
condition $c_+=0$, \eq{nrr} reduces to a two-term recursion relation.

The third normalizable canonical form has $P(z)=\nu z^2$, and therefore
$c_{+0}=c_{0-}=0$. The exactly solvable potentials are characterized by the
vanishing of the coefficients $c_+$ or $c_-$, but not both simultaneously.
In the former case we can take $w=z$, while in the latter $w$
is proportional to $1/z$ (see the foregoing discussion on the vanishing of the
critical coefficient \eq{ccoeff}). In either case, the coefficient of
$P_{k-1}$ in \eq{nrr} vanishes identically.

The fourth normalizable canonical form is given by $P(z)=z$, so that
$w=z$ and $\c_{+0}=c_{+0}=0$, and its exactly solvable potentials are defined
by the vanishing of the coefficient $c_+=\c_+=0$, so that \eq{nrr} is
two-term. Finally, for the fifth normalizable canonical form $P(z)=1$ all
normalizable potentials are automatically exactly solvable (they are
translates of the harmonic oscillator), and we have already seen that its
associated orthogonal polynomials satisfy the two-term recursion relations
\eq{reo}.
\qed

\Section{op} The Orthogonal Polynomials.

We shall study in this section the properties of the family of weakly
orthogonal polynomials associated to a \qes. one-dimensional
Hamiltonian in the manner described in the previous section. 
Since, as we shall see, these properties can be established directly from the
recursion relation \eq{nrr} or
\eq{reo}, these polynomials have basically the same properties as
those studied by Bender and Dunne in \rf{BenDun}.

We have seen in the previous section that the polynomials $\Phat(E)$
defined by a \qes. one-dimensional Hamiltonian satisfy a three-term recursion
relation of the form
$$
\Phat_{k+1} = (E-b_k)\,\Phat_k-a_k\,\Phat_{k-1},\qquad k\ge0,
\Eq{crr}
$$
with $a_0=0$ and
$$
a_{n+1} = 0.
\Eq{an1}
$$
For the fifth canonical form, the polynomials $P^0_k$ and $P^1_k$ also
satisfy a recursion relation of the form \eq{crr}, with $a_k=0$ for all
$k\ge0$. Note that the coefficients $a_k$, $b_k$ in \eq{crr} are
guaranteed to be real only for the canonical forms 2--7 and 10 (for which $P$
has a real root). As remarked in the previous section, the vanishing of $a_k$
for a positive integer value of $k$ means that the polynomials $\Phat_k$ are
only weakly orthogonal. In particular, many classical results, based on the
fact that $a_k>0$ (or sometimes $a_k\ge0$) for $k\ge 1$ cannot be applied in
our case.

By Favard's theorem, \rf{Chi}, there is a {\it moment
functional}, that is a linear functional $\CL$ acting in the space $\C[E]$ of
(complex) univariate polynomials, such that the polynomials $\Phat_k$ are
orthogonal under $\CL$:
$$
\CL(\Phat_k\,\Phat_l) = \gamma_k\,\delta_{kl},\qquad k,l\in\Nn.
\Eq{orth}
$$
The functional $\CL$ is unique if we impose the normalization condition
$\CL(\Phat_0)=\CL(1)=1$.
It is also known (Boas's theorem, \rf{Chi}) that there is a (not
necessarily unique) function of bounded variation $\omega$ such that 
$$\CL(p)=\int_{-\infty}^{\infty}p(E)\,d\omega(E)
\Eq{omega}$$
for an arbitrary polynomial $p$. The coefficient $\gamma_k=\CL(\Phat_k^2)$,
which therefore plays the role of the square of the norm of $\Phat_k$, can be
computed by multiplying \eq{crr} by $\Phat_{k-1}$ and taking $\CL$ of both
sides, obtaining
$$
0 = \gamma_k-a_k\gamma_{k-1},\qquad k\ge 1.
$$
Taking into account that $\gamma_0=\CL(1)=1$ we get
$$
\gamma_k = \prod_{j=1}^k a_j,\qquad k\ge1.
\Eq{gamma}
$$
In particular, from this formula follows one of the key properties of the
weakly orthogonal polynomial system associated to a one-dimensional \qes.
Hamiltonian. Namely, from \eq{an1} we have
$$
\gamma_k=0,\qquad k\ge n+1,
$$
so that {\it all the polynomials $\Phat_k$ with $k\ge n+1$ have zero norm.}
From this formula it also follows that the ``squared norms" $\gamma_k$ will be
positive for $k\le n$ if and only if $a_k>0$ for $1\le k\le n$. It can be shown
by a straightforward computation that this is always the case when $P$ is one
of canonical forms 2--4 in \eq{normlist}, assuming that all the eigenfunctions
of $H$ are square-integrable and that $H$ is {\it not} exactly solvable. Note
also that when $H$ is normalizable (canonical forms 1--5 in \eq{normlist}) and
exactly solvable then
$a_k=0$ for all $k\ge0$. Hence the square norms of all the polynomials
$\Phat_k$ vanish, from which it easily follows from \eq{crr} that
$\CL=\delta(E-b_0)$.

Other important properties of the polynomials $\Phat_k$ concern their zeros.
Classically, \rf{Chi}, it can be shown that if $a_k>0$ for all $k\in\Nn$ then
the zeros of the polynomials $\Phat_k$ satisfying a three-term recursion
relation
\eq{crr} are real and simple. In our case the condition $a_k>0$ for all
$k\in\Nn$ can never hold on account of \eq{an1}. However, if $H$ is fully
algebraic it can still be proved that all the zeros of $\Phat_{n+1}$ are real
and simple.  Indeed, by hypothesis $H$ is self-adjoint on the space $\N$ of
functions of the form \eq{algwf}, with $\chi\in\CP_n$. Hence $H$ has $n+1$
linearly independent algebraic eigenfunctions lying in $\N$, whose
corresponding eigenvalues are real (by self-adjointness) and distinct ($H$
being a one-dimensional Sturm--Liouville operator). Let us denote by $E_0<
E_1<\dots< E_n$ these $n+1$ real eigenvalues of $H$ on $\N$, and by
$\psi_l(x)\equiv\psi_{E_l}(x)$ the eigenfunction corresponding to the
eigenvalue
$E_l$. Then \eq{algwf} and either \eq{xalt1} or \eq{xalt2} imply that
$P_k(E_l)=0$, or equivalently $\Phat_k(E_l)=0$, for $k\ge n+1$ and $0\le l\le
n$. In particular, since $\Phat_{n+1}$ is of degree $n+1$ and all the
eigenvalues $E_l$ are different, it follows that
$$
\Phat_{n+1}(E) = \prod_{l=0}^n(E-E_l),
\Eq{pn1}
$$
where we have used \eq{crr} and the fact that $\Phat_0=1$. In other words, {\it
$\Phat_{n+1}$ has $n+1$ simple real zeros at the $n+1$
algebraic eigenvalues of $H$.}
Furthermore, from the fact that
$\Phat_{k}$ vanishes at $E_l$ for $k\ge n+1$ we conclude that there exist monic
polynomials $Q_k$ of degree $k$ such that
$$
\Phat_{k+n+1} = Q_k \Phat_{n+1},\qquad k\ge 0.
\Eq{fp}
$$
This is the so called {\it factorization property} of the polynomial system
$\{\Phat_k\}_{k\in\Nn}$, \cf \rf{BenDun}. Note that the vanishing of
$\Phat_k(E_l)$ for all $k\ge n+1$ is consistent with the recursion relation on
account of \eq{an1}. In fact, when $a_k$ is positive for $k\ge1$ and $b_k$ is
real for $k\ge0$, \eq{pn1} follows directly from the recursion relation by
\lm{lopos}, without using the fact that the polynomials $\Phat_k$ are associated to a
fully algebraic \qes. one-dimensional Hamiltonian. The vanishing of
$\Phat_k(E_l)$ for $k>n+1$ is then an immediate consequence of
$\Phat_{n+1}(E_l)=0$, the recursion relation \eq{crr} and \eq{an1}.

From the previous equation and \eq{crr} it follows that the
polynomials $Q_k$ also satisfy a three-term recursion, namely
$$
Q_{k+1} = (E-b_{k+n+1}) Q_k - a_{k+n+1} Q_{k-1},\qquad k\ge0,
$$
and are therefore orthogonal with respect to an appropriate moment functional
$\CL_Q$ (in general different from $\CL$).

It was heuristically argued in \rf{KUW} that
$$
\CL = \sum_{j=0}^n \omega_j\,\delta(E-E_j)
\Eq{mf}
$$
on $\C[E]$, where the coefficients $\omega_j$ are defined by
$$
\sum_{l=0}^n \Phat_k(E_l)\,\omega_l = \delta_{k0},\qquad k=0,1,\dots,n.
\Eq{os}
$$
Equivalently, the discrete Stieltjes measure
$d\wh(E)$ defined by the function
$$
\wh(E) = \sum_{j=0}^n \omega_j\,\theta(E-E_j),
\Eq{wd}
$$
where $\theta(t)$ is Heaviside's step function, satisfies \eq{omega}.
Note that the linear system \eq{os} uniquely defines the $n+1$ constants
$\omega_j$, since by \eq{alt1} or \eq{alt2} its coefficient matrix is the
matrix of the change of basis
$\left\{c_k(z-z_1)^k/k!\right\}_{k=0}^n$ or
$\left\{c_k(z-z_1)^k(z-z_2)^{n-k}/k!\right\}_{k=0}^n$
to $\left\{\chi_{E_l}\right\}_{l=0}^n$ in $\C\otimes\CP_n$, $c_k$ being the
coefficient of $\Phat_k$ in \eq{phatk}. It is not difficult to show rigorously
that
\eq{orth} is satisfied. Indeed, by the uniqueness of $\CL$ this is
equivalent to showing that if $\CL_0 =
\sum_{j=0}^n \omega_j\,\delta(E-E_j)$ then
$$
\CL_0(\Phat_k\Phat_l) = 0,\qquad k\ne l,
\Eq{orthomeg}
$$
and that
$$
\CL_0(\Phat_0) = \CL_0(1) = 1,
$$
since $\CL(\Phat_k^2)$ and $\CL_0(
\Phat_k^2)$ must coincide if \eq{orthomeg} holds due to the recursion
relation \eq{crr}. From the definition of $\omega_j$ we deduce that the last
equation, together with \eq{orthomeg} for $k=0$ and $l=1,\dots,n$, are
satisfied. Suppose now that \eq{orthomeg} holds for $k=0,1,\dots,K$
($K\le n-1$) and
$k<l\le n$. Multiplying \eq{crr} by $\Phat_l$ and taking $\CL_0$ of both sides we
obtain
$$
\CL_0(\Phat_{K+1}\Phat_l) = \CL_0((E-b_K)\Phat_{K}\Phat_l)-a_K\CL_0(\Phat_{K-1}\Phat_l) =
\CL_0(E\Phat_K\Phat_l)
$$
if $K+1<l\le n$, by the induction hypothesis. But, using again \eq{crr},
$$
\CL_0(E\Phat_K\Phat_l) = \CL_0(\Phat_K\cdot E \Phat_l) = \CL_0(\Phat_K\Phat_{l+1})+
b_l\CL_0(\Phat_K\Phat_l) + a_l \CL_0(\Phat_K\Phat_{l-1}) = 0,
$$
by the induction hypothesis (since $l>K+1$ implies $l-1>K$). Hence
\eq{orthomeg} is true for
$0\le k,l\le n$. Finally, \eq{orthomeg} is trivially true when $k$ or $l$ are
greater than $n$ by the factorization property \eq{fp} and \eq{pn1}.

We shall next show that all the coefficients $\omega_j$ are positive if
$b_k$ is real for all $0\le k\le n$ and $a_k>0$ for $1\le k\le n$. (Several
instances of this result were checked numerically in
\rf{KUW} for the orthogonal polynomials associated to the Hamiltonian \eq H.)
The proof is based on the following simple lemma:
\Lm{lopos} 
If $a_k>0$ for $k=1,2,\dots, n$ and $b_k$ is real for $k=0,1,\dots,n$ then
$\CL$ is {\rm positive-definite} on
$\CP_{2n}$. In other words, if $p\in\CP_{2n}$ is a real polynomial of degree
at most $2n$, $p\ne0$ and $p(E)\ge0$ for all $E\in\R$ then $\CL(p)>0$.

\Proof
A polynomial $p\in\CP_{2n}$ which is non-negative for all real values of $E$
must be of the form $q^2+r^2$, where $q,r\in\CP_n$ are real polynomials.
Write $q=\sum_{k=0}^l q_k \Phat_k$; then all the coefficients $q_k$ are
real, since $\Phat_k$ is a real polynomial for $0\le k\le n$ by the hypotheses.
Using the orthogonality of the polynomials $\Phat_k$ we obtain
$\CL(q^2) = \sum_{k=0}^n q_k^2 \gamma_k$. Similarly, if $r=\sum_{k=0}^l
p_k\Phat_k$ then $\CL(r) = \sum_{k=0}^n r_k^2 \gamma_k$, and
$\CL(p)=\sum_{k=0}^n (q_k^2+r_k^2) \gamma_k$. Since $\gamma_k>0$ for
$k=0,1,\dots,n$ by \eq{gamma} and the hypothesis on the coefficients $a_k$, it
follows that
$\CL(p)\ge0$, and $\CL(p)=0$ if and only if $q_k=r_k=0$ for $k=0,1,\dots,n$,
that is if $p=0$.\qed

\Pr{opos}
If $a_k>0$ for $k=1,2,\dots, n$ and $b_k$ is real for $k=0,1,\dots,n$ then
$\omega_k>0$ for all $k=0,1,\dots n$.

\Proof
Apply the previous lemma to the polynomials
$\prod_{0\le j\ne k\le n}(E-E_j)^2\in\CP_{2n}$ for $k=0,1,\dots n$.\qed

Note that the hypotheses of the previous proposition are satisfied when $P$
is one of canonical forms 2, 3 or 4, provided that all the eigenfunctions
of $H$ are square-integrable and that $H$ is {\it not}\/ exactly solvable. In
particular, it is satisfied by the Hamiltonian \eq{H}.

The (Hamburger) {\it moment problem} for the moment functional \eq{mf}
associated to the weakly orthogonal polynomials defined by a \qes.
one-dimensional Hamiltonian consists in determining whether there is a {\it
distribution function} (\ie a non-decreasing
function of bounded variation)
$\omega$ such that $\CL$ can be represented by \eq{omega} 
for an arbitrary polynomial $p$. We have already shown that this problem
has a solution \eq{wd}, since \eq{wd} is clearly non-decreasing and of bounded
variation. We shall next show that this solution is unique (up to
an additive constant), so that the moment problem associated to
the weakly orthogonal polynomial system $\{\Phat_k\}_{k\in\Nn}$ is
always determined \rf{fnote3}. Essentially, this is due to the
fact that the {\it spectrum}
$$
\sigma(\wh) =
\Set{E\in\R}{\wh(E+\delta)-\wh(E-\delta)>0,\;\forall\delta>0}
$$
of the distribution function \eq{wd} is the finite set
$\left\{E_l\right\}_{l=0}^n$ \rf{fnote4}. According to a well known result in
the classical theory of orthogonal polynomials, \rf{Chi}, a distribution
function
$\omega$ defines a positive-definite functional on $\C[E]$ through integration
with respect to the Stieltjes measure $d\omega(E)$ if and only if the spectrum of
$\omega$ is infinite. Since $\CL$ is not positive-definite
($\CL(\Phat_{n+1}^2)=\gamma_{n+1}=0$), any solution
$\omega$ of \eq{omega} must have a finite spectrum, and will thus be of the form
$$
\omega(E) = \sum_{k=0}^{\tilde n} \tilde \omega_k \theta(E-\Etilde_k)+C
$$
for some constant $C$, up to an immaterial redefinition of $\omega$ in
$\sigma(\omega)$. If $I$ is a compact interval containing
$\sigma(\wh)\union\sigma(\omega)$, then
$$
\CL(p) = \int_I p(E)\,d\wh(E) = \int_I p(E)\,d\omega(E),\qquad\forall
p\in\C[E].
$$
Since $I$ is compact, a well known theorem (\cf \rf{Chi}) shows that $\wh$
and $\omega$ differ by a constant at all points in which both $\wh$ and
$\omega$ are continuous. But this easily implies that
$E_k=\Etilde_k$ and $\omega_k = \tilde \omega_k$ for $k=0,1,\dots,n=\tilde n$,
whence
$\omega=\wh+C$, as stated. Note that the same argument shows that the moment
problem in any interval containing $[E_0,E_n]$; in particular, the
(Stieltjes) moment problem in $[E_0,\infty)$ is also determined. In this
respect, the weakly orthogonal polynomials associated to a \qes.
one-dimensional Hamiltonian behave in exactly the same way as the classical
orthogonal polynomials, whose moment problem
is also determined,
\rf{Chi}.

The {\it moments} of the moment functional $\CL$ are by definition the
numbers
$$
\mu_k = \CL(E^k) = \int_{-\infty}^\infty E^k\,d\wh(E) =
\sum_{l=0}^n \omega_l\,E^k_l,\qquad k\in\Nn.
\Eq{muk}
$$
If the hypotheses of \pr{opos} hold, all the moments are real. From \eq{muk}
we see that the module of the $k$-th moment $\mu_k$
does not grow factorially as $k$ tends to infinity, as argued in \rf{BenDun},
but instead it diverges like the $k$-th power of a constant \rf{fnote5}.

We shall next show that if the coefficient $a_k$ satisfies the condition
$$
a_k\ne0,\qquad 1\le k\le n,
\Eq{ahp}
$$
which guarantees that the polynomials $\Phat_k$ have non-zero norm for $k\le
n$, then {\it the moments $\mu_k$ with $k\ge n+1$ satisfy a constant
coefficient difference equation of order $n+1$.}
To this end, recall first of all that the
bilinear form $\inner{p,q} = \CL(p\,q)$ defined by $\CL$ in $\C[E]$, when
restricted to the subspace $\C\otimes\CP_l$, is represented in the basis
$\{E^k\}_{0\le k\le l}$ by the symmetric matrix
$(\mu_{i+j})_{0\le i,j\le l}$, whose determinant we shall denote by
$\Delta_l$. On the other hand, the matrix of the bilinear form
$\inner{\cdot\,,\,\cdot}$ in the basis $\{\Phat_k\}_{0\le k\le l}$
is clearly
$\diag(1,\gamma_1,\dots,\gamma_l)$; therefore, by \eq{gamma} and the hypothesis
on the coefficients $a_k$, we conclude that $\Delta_n\ne0$ and
$$
\Delta_{k} = 0,\qquad k\ge n+1.
\Eq{dz}
$$
In particular, since $\Delta_n\ne0$ but $\Delta_{n+1}=0$, the last column of
$\Delta_{n+1}$ must be a linear combination of the remaining columns, so that
$$
\mu_k = \sum_{i=1}^{n+1} c_i\,\mu_{k-i},\qquad n+1\le k\le 2(n+1),
\Eq{murr}
$$
for some (in general complex) constants $c_1,\dots,c_{n+1}$. An easy
induction argument using \eq{dz} then shows that the above relation is
actually valid with the {\it same} constant coefficients $c_i$ for all $k\ge
n+1$, as claimed. In fact, it is not hard to see that $c_i$ in
\eq{murr} is minus the coefficient of $E^{n+1-i}$ in $\Phat_{n+1}$. Indeed,
write
$\Phat_{n+1}=E^{n+1}-p_n$, with
$$
p_n = \sum_{i=1}^{n+1} \tilde c_i\,E^{n+1-i},
$$
and let $Q_k = E^k-q_{k-1}$, so that $q_{-1}=0$ and $\deg q_{k-1}\le k-1$ for
$k\ge 1$. From
\eq{fp} it follows that
$$
\Phat_{k} = E^{k}-E^{k-n-1} p_n - q_{k-n-2}\Phat_{n+1},\qquad k\ge n+1,
$$
which by \eq{orth} implies that
$$
\mu_{k} = \CL(E^{k}) = \CL(E^{k-n-1}p_n) =
\sum_{i=1}^{n+1} \tilde c_i\,\mu_{k-i},\qquad n+1\le k\le 2(n+1).
$$
Comparing with \eq{murr} and taking into account the linear independence of the
columns of $\Delta_n$ we immediately obtain that $\tilde c_i = c_i$ for
$i=1,2,\dots,n+1$, as stated.

Note that the fact that the moments satisfy a constant coefficient recursion
relation \eq{murr} (with $k\ge n+1$) actually characterizes weakly orthogonal
polynomial systems. Indeed, \eq{murr} simply expresses the fact that the
$(n+2)$-th column of $\Delta_l$ for $l\ge n+1$ is a linear combination of the
first $n+1$ columns. Hence the recursion relation implies \eq{dz}, and since
$\Delta_{n+1} = \prod_{j=1}^{n+1} \gamma_j$ this means that $\gamma_k=0$ for
some $k\le n+1$, so that $a_k=0$ for some $k\le n+1$ by \eq{gamma}.

Consider, for example, the Hamiltonian \eq{H} studied in \rf{BenDun}, which
corresponds to the fourth canonical form with
$$
n = J-1,\quad c_+=-16,\quad c_0=c_*=0,\quad c_-=2s+\frac12(n-1).
\Eq{hcoeffs}
$$
The coefficients of the corresponding recursion relation \eq{crr} are easily
found to be
$$
b_k=0,\quad a_k = 16k(J-k)(k+2s-1),\qquad k\ge0.
\Eq{hrr}
$$
Since we can take $s\ge1/2$ without loss of generality, we see that $a_k>0$
for $1\le k\le n$, so that \eq{ahp} is satisfied. Furthermore, since $b_k$
vanishes for all
$k\ge0$ the polynomials
$\Phat_k$ have parity
$(-1)^k$, and therefore all the odd moments vanish (the corresponding moment
functional is said to be {\it symmetric}). For
$J=3$ (that is, $n=2$), according to the foregoing observations we know that
the moments satisfy a third-order recursion relation of
the form \eq{murr}, whose coefficients are minus the coefficients of $E^2$, $E$
and
$1$ in $\Phat_3$. From \eq{crr} (with $\Phat_0=1$) we obtain
$$
\Phat_1 = E,\qquad \Phat_2=E^2-64s,\qquad \Phat_3(E)=E^3-32(4s+1)E,
\Eq{ps}
$$
so that $c_1=c_3=0$---as expected, since the moment functional is
symmetric---, and
$c_2=32(4s+1)$.
Therefore the even moments satisfy the first-order recursion relation
$$
\mu_{2j} = 32(4s+1) \mu_{2j-2},\qquad j\ge2,
\Eq{mu2j}
$$
and since
$\mu_2 = \gamma_1 = a_1 = 64s$, from \eq{mu2j} we obtain
$$
\mu_{2j} = 32^{j-1}(4s+1)^{j-1}\cdot 64s,\qquad j\ge1.
\Eq{mu2js}
$$
Thus, in this case $\mu_{2j}$ has a pure power growth. The same result
can be obtained using \eq{muk}. Indeed, from
\eq{ps} we have
$$
E_0=-\lambda\equiv-\sqrt{32(4s+1)},\qquad E_1=0,\qquad E_2=\lambda,
$$
and therefore
$$
\omega_0=\frac
s{4s+1},\qquad\omega_1=\frac{2s+1}{4s+1},\qquad\omega_2=\omega_0
$$
from \eq{os} and \eq{ps}.
Thus
$$
\mu_k = \frac s{4s+1}\left[(-\lambda)^k+\lambda^k\right],
$$
which yields $\mu_{2j+1}=0$ for $j\ge0$ and \eq{mu2js}.

\Section{conc} Conclusions.

We have shown in this paper how every \qes. one-dimensional Hamiltonian
satisfying conditions \eq{c1}--\eq{c9} defines a weakly orthogonal polynomial
system $\{\Phat_k\}_{k=0}^\infty$ through the three-term recursion relation
\eq{nrr} (with initial condition $\Phat_0=1$). It is important, in this
context, to emphasize the {\it weak} orthogonality of the polynomials
$\Phat_k$, \ie the fact that the norm of $\Phat_k$ may vanish---and in fact
{\it does} vanish for $k\ge n+1$, $n$ being the ``spin" parameter present in
the Hamiltonian. As explained in \section{op}, this is an inevitable
consequence of the vanishing of the coefficient of $\Phat_{k-1}$ in the
recursion relation \eq{nrr} for $k=n+1$, which is made possible by the fact
that the parameter $n$ is a non-negative integer. The latter fact, however, is
an intrinsic property of one-dimensional \qes. (as opposed to merely
Lie-algebraic) Hamiltonians; indeed, it is a key factor in the explanation of
the partial integrability of a \qes. Hamiltonian outlined in \section{qes}. To
better illustrate this point, consider the Hamiltonian \eq{H}, which is
Lie-algebraic for all real values of the parameter $J$. Indeed, $H$ can be
written in the form
\eq{hgh}--\eq{hgj}, with $\zeta(x)=x^2/4$, $\mu(z) = e^{-4z^2}z^{s-1/4}$,
$c_{++}=c_{+0}=c_{00}=c_{+-}=c_{--}=0$, $c_{0-}=1/2$, and the
remaining coefficients given by \eq{hcoeffs}, where now $n$ is to be regarded
as an arbitrary real parameter. When $n$ is not a non-negative integer, the
generators
\eq{jeps} don't leave invariant any finite-dimensional polynomial module
$\CP_n$, so that $H$ is in general non-integrable---there is no special
reason for $H$ to have algebraically computable eigenfunctions of the form
\eq{algwf}, with $\chi$ a polynomial. However, even when $n$ is not
a non-negative integer, the Lie-algebraic nature of $H$ and conditions
\eq{cond} imply that the polynomials $\Phat_k$ defined by \eq{algwf},
\eq{chie} and \eq{phatk} still satisfy a three-term recursion relation
\eq{crr}, with the coefficients given by \eq{hrr}. 
In other words, what makes $H$ \qes. is not
merely the fact that its associated polynomials satisfy a three-term recursion
relation \eq{crr} (which implies their orthogonality with respect to
some Stieltjes measure), but the fact that the coefficient
$a_k$ in this recursion relation vanishes for some positive integer value of
$k$, so that the associated polynomials $\Phat_k$ can only be weakly
orthogonal.

As we saw in \section{op}, the Stieltjes measure with respect to which the
polynomials $\Phat_k$ associated to a \qes. Hamiltonian $H$ are orthogonal is
supported in the set of algebraic eigenvalues of $H$, which is a finite set.
For this reason, the polynomials $\Phat_k$ are {\it discrete
polynomials.} Although the classical (Hermite, Legendre, Laguerre, Tchebycheff,
etc.) polynomials of Mathematical Physics are orthogonal with respect to a
continuous measure, discrete (Charlier, Hahn, Krawtchouk, Meixner,
Tchebycheff, etc.) polynomials have also been studied in the mathematical
literature of orthogonal polynomials, \cf\rf{Chi}. Note that a discrete
polynomial system is truly---as opposed to weakly---orthogonal if and only if
the supporting set of its Stieltjes measure is infinite. Some of the discrete
polynomials cited above, like the Hahn, Krawtchouk or discrete Tchebycheff
polynomials, are in fact weakly orthogonal. In general, weakly
orthogonal polynomials arise naturally, for instance, in the theory of
approximate polynomial curve fitting, \rf{Pec}. More recently, \rf{SmiTur}, the
study of second-order finite difference eigenvalue equations with infinitely
many polynomial solutions has led to an interesting connection between a
non-standard finite-dimensional representation of
$\sL2$ and certain families of weakly orthogonal discrete polynomials
(Hahn polynomials and analytically continued Hahn polynomials).

Let us stress, in closing, that the present paper deals only with
one-dimensional \qes. Hamiltonians. It is an interesting open problem to
generalize these results to \qes. multi-dimensional systems, a possibility
already considered in \rf{KUW}, where a heuristic (but inconclusive,
in our opinion) argument was advanced suggesting that all \qes. systems give
rise to weakly orthogonal polynomials. In the two-dimensional
case, at least, the classification of \qes. Lie algebras of first-order
differential operators in two variables presented in
\rf{GKOqes} and \rf{GKOreal} could be used as a starting point for an analysis
along the present lines.

\Section{ack} Acknowledgments.
It is a pleasure to thank M. A. Mart\'\i n-Delgado, who
pointed out to us reference \rf{BenDun}, Gabriel \'Alvarez, for useful
discussions regarding the theory of classical orthogonal polynomials, and
Carlos Finkel, for providing several key references.

The authors would also like to acknowledge the partial financial support of the
DGICYT under grant no.~PB92--0197.

\bigskip\bigskip
\References
\bye